\documentclass[12pt]{article}
\usepackage{amsmath}
\usepackage{amssymb}
\usepackage{amsfonts}
\usepackage{latexsym}
\usepackage{color}

\catcode `\@=11 \@addtoreset{equation}{section}
\def\theequation{\arabic{section}.\arabic{equation}}
\catcode `\@=12

\newcommand{\be}{\begin{equation}}
\newcommand{\en}{\end{equation}}
\newcommand{\bea}{\begin{eqnarray}}
\newcommand{\ena}{\end{eqnarray}}
\newcommand{\beano}{\begin{eqnarray*}}
\newcommand{\enano}{\end{eqnarray*}}
\newcommand{\bee}{\begin{enumerate}}
\newcommand{\ene}{\end{enumerate}}

\newcommand{\N}{N \!\!\!\!\! N}

\newcommand{\Hil}{{\cal H}}

\newcommand{\Id}{1\!\!1}

\newcommand{\Lc}{{\cal L}}
\newcommand{\D}{{\cal D}}
\newcommand{\C}{{\cal C}}
\newcommand{\E}{{\cal E}}
\newcommand{\Nc}{{\cal N}}
\newcommand{\M}{{\cal M}}

\newtheorem{thm}{Theorem}

\newtheorem{prop}[thm]{Proposition}

\newenvironment{proof}{\noindent {\bf Proof:}}{\hfill$\Box$}

\begin{document}

\thispagestyle{empty}

\vspace*{1cm}

\begin{center}
{\Large \bf An invariant analytic orthonormalization procedure with an application to coherent states}   \vspace{2cm}\\

{\large F. Bagarello}
\vspace{3mm}\\
  Dipartimento di Metodi e Modelli Matematici,
Fac. Ingegneria, Universit\`a di Palermo, I-90128  Palermo, Italy\\
e-mail: bagarell@unipa.it
\vspace{2mm}\\
{\large S. Triolo}
\vspace{3mm}\\
  Dipartimento di Matematica ed Applicazioni,
 Universit\`a di Palermo, I-90128  Palermo, Italy\\
e-mail: salvo@math.unipa.it
\end{center}

\vspace*{2cm}

\begin{abstract}
\noindent We discuss a general strategy which produces an
orthonormal set of vectors, stable under the action of a given set
of unitary operators $A_j$, $j=1,2,\ldots,n$, starting from a
 fixed normalized vector in $\Hil$ and from a set of unitary operators. We
discuss several examples of this procedure and, in particular, we
show how a set of {\em coherent-like} vectors can be produced and in
which condition over the lattice spacing this can be done.
\end{abstract}

\vspace{2cm}


\vfill

\newpage

\section{Introduction}

In the mathematical and physical literature many examples of
complete sets of vectors in a given Hilbert space $\Hil$ are
constructed starting from a single normalized element $f_0\in\Hil$,
 acting on this vector several time with a given set of unitary
operators. As a matter of fact, this is exactly what happens for
coherent states and for wavelets, just to cite maybe the most known
examples. In the first case one essentially acts several times on
the vacuum of a bosonic oscillator with a modulation and a
translation. In the second example, to produce a complete set of
wavelets one acts respectively  on a {\em mother wavelet} with
powers of  a dilation and a translation operator. In this last
situation the result of this action can be an o.n. set of vectors,
and this is the main result of the so-called multi-resolution
analysis, \cite{dau}, while this is forbidden for general reasons
for coherent states. Both these examples, as well as many others,
can be considered as particular cases of a general procedure in
which a certain set of vectors is constructed acting on a fixed
element of $\Hil$, $f_0$, with a certain set of unitary operators,
$A_1,\ldots,A_N$: $f_{k_1,\dots,k_N}:=A_1^{k_1}\cdots A_N^{k_N}f_0$,
$k_j\in\Bbb{Z}$ for all $j=1,2,\ldots,n$. These vectors may or may
not be orthogonal: we consider here the problem of {\em
orthonormalizing this set}, i.e. the problem of producing a new set
of vectors which share with the original one most of its features
and, moreover, are also orthonormal.

The paper is organized as follows:

in the next section we state the general problem, discuss the
method and show some {\em prototype} examples.

In Section III we discuss in many details the example concerning the
coherent states, and we find conditions for our orthonormalization
procedure to work. In particular we show that, under certain
conditions on a parameter which can be interpreted as a
two-dimensional lattice spacing, a set of vectors can be obtained
which shares with the coherent states a number of properties. To be
explicit this new set satisfies indeed a closure condition in a
certain Hilbert space, is an o.n. set of vectors, and is stable
under the action of the same unitary operators which generate the
set of coherent states. Moreover, each element of this new set is an
eigenstate of a annihilation-like operator and saturates the
Heisenberg uncertainty relation.

Section IV contains our final considerations and  plans for the
future.

The paper ends with an Appendix on a generalized version of the
$(k,q)-$representation which is widely used in Section III.

\section{Stating the problem and first results}

Let $\Hil$ be a Hilbert space, $f_0\in\Hil$ a fixed element of the
space and $A_1,\ldots,A_N$ $N$ given unitary operators:
$A_j^{-1}=A_j^\dagger$, $j=1,2,\ldots,N$. Let $\Hil_N$ be the
closure of the linear span of the set
\be\Nc_N=\{f_{k_1,\ldots,k_N}:=A_1^{k_1}\cdots A_N^{k_N}f_0,\,\,
k_1,\ldots k_N\in\Bbb{Z} \}\label{II1}\en Of course, in order for
this situation to be of some interest, it is necessary to assume
that the vectors in $\Nc_N$, or part of them, are linearly
independent: indeed, if this is not the case we may likely get an
Hilbert space $\Hil_N$ which has finite dimension, and this is
something not very interesting for us. Therefore in the following
we will assume that all the vectors $f_{k_1,\ldots,k_N}$ are
independent and it is clear, by the definition itself, that they
are also complete in $\Hil_N$. In general there is no reason why
the vectors in $\Nc_N$ should be mutually orthogonal. On the
contrary, without a rather clever choice of both $f_0$ and
$A_1,\ldots,A_N$, it is very unlikely to obtain an o.n. set. Our
aim is to discuss some general technique which produces another
vector $\varphi\in\Hil_N$ such that the set
\be\M_N=\{\varphi_{k_1,\ldots,k_N}:=A_1^{k_1}\cdots
A_N^{k_N}\varphi,\,\, k_1,\ldots k_N\in\Bbb{Z} \}\label{II2}\en is
made of orthogonal vectors. Moreover, we would like this set to
share as much of the  original features of $\Nc_N$ as possible.
For instance, if the set $\Nc_N$ is a set of coherent states, we
would like the new vectors $\varphi_{k_1,\ldots,k_N}$ to be, for
instance, eigenstates of a (sort of) annihilation operator, to
give rise to a resolution of the identity and to saturate the
Heisenberg uncertainty relation.

We will analyze this problem step by step, starting with the
simplest situation which is, clearly, $N=1$. In this case the set
$\Nc_1$ in (\ref{II1}) reduces to $\Nc_1 =\{f_{k}:=A^{k}\,f_0,\,\,
k\in\Bbb{Z} \}$ with $<f_k,f_l>\neq \delta_{k,l}$ (otherwise we
have already solved the problem!). Since $\Nc_1$ is complete in
$\Hil_1$, any element in $\Hil_1$ can be written in terms of the
vectors of $\Nc_1$. Let $\varphi_0\in\Hil_1$ be the following
linear combination:
\be\varphi_0=\sum_{k\in\Bbb{Z}}\,c_kf_k,\label{II3}\en and let us
define more vectors of $\Hil_1$ as
\be\varphi_n=A^n\varphi_0=\sum_{k\in\Bbb{Z}}\,c_k\,f_{k+n}=Xf_{n},\label{II4}\en
where we have introduced the  operator \be
X=\sum_{k\in\Bbb{Z}}\,c_k\,A^k.\label{II5}\en The coefficients
$c_k$ should be fixed by the following orthogonalization
requirement: $<\varphi_n,\varphi_0>=\delta_{n,0}$. It is worth
remarking that all the expansions above are, for the moment, only
formal. What makes everything well defined is the asymptotic
behavior of the coefficients of the expansion $c_n$, and we will
discuss in the rest of the paper, and in particular in section
III, that there exist situations in which the series for
$\varphi_n$ and $X$ do converge and other situations in which they
do not.

The first useful result is  that if
$<\varphi_n,\varphi_0>=\delta_{n,0}$ for all $n\in\Bbb{Z}$, then
$<\varphi_n,\varphi_k>=\delta_{n,k}$, $\forall n,k\in\Bbb{Z}$. This
follows directly from the definition of $\varphi_n$ since
$$
<\varphi_n,\varphi_k>=<A^n\varphi_0,A^k\varphi_0>=<A^{n-k}\varphi_0,\varphi_0>=<\varphi_{n-k},\varphi_0>=
\delta_{n-k,0}.
$$
For this reason, in order to fix the coefficients $c_n$, it is
enough to require the orthogonality condition
$<\varphi_n,\varphi_0>=\delta_{n,0}$, which becomes \be
\delta_{n,0}=<\varphi_n,\varphi_0>=\sum_{k,l\in\Bbb{Z}}\,\overline{c_k}\,c_l\,<f_{k+n},f_l>=
\sum_{k,l\in\Bbb{Z}}\,\overline{c_k}\,c_l\,a_{k+n-l},
\label{II6}\en where we have defined \be
a_j=<A^jf_0,f_0>\label{II7}\en If we now multiply both sides of
(\ref{II6}) for $e^{ipn}$ and  sum up on $n\in\Bbb{Z}$ we get \be
|C(p)|^2\,\alpha(p)=1,\quad \mbox{a.e. in }
[0,2\pi[,\label{II8}\en where we have introduced the following
functions:\be C(p)=\sum_{l\in\Bbb{Z}}\,c_l\,e^{ipl},\quad
\alpha(p)=\sum_{l\in\Bbb{Z}}\,a_l\,e^{ipl}\label{II9} \en Again,
these series are not necessarily convergent, so that they must be
considered only as formal objects at this stage.

In particular, it is an easy exercise to check that, if the
following quantities all exist, then
$\sum_{l\in\Bbb{Z}}\,|c_l|^2=\frac{1}{2\pi}\int_0^{2\pi}|C(p)|^2\,dp=\frac{1}{2\pi}\int_0^{2\pi}\frac{dp}{\alpha(p)}$.
This result suggests that for particular choices of $f_0$ and $A$ it
might happen that the series for $\alpha(p)$ is not convergent or,
even if it converges to a $2\pi$-periodic and $C^\infty$ function,
this function might have  in $[0,2\pi[$ a zero which makes of
$\alpha(p)^{-1}$ a nonintegrable function. If this is the case there
is no reason to claim that the sequence $\{c_l\}$ belongs to
$l^2(\Bbb{Z})$. On the contrary, any time that the function
$\alpha(p)$ exists as a continuous function, i.e. under suitable
conditions on the $a_l\,'s$ which are satisfied in many relevant
situations, e.g. for coherent states, and if $\alpha(p)$ does not
vanish in $[0,2\pi[$, we can conclude that $\{c_l\}\in
l^2(\Bbb{Z})$. But, in this case, we can do much better than this:
since $\{c_l\}\in l^2(\Bbb{Z})$ then $C(p)\in\Lc^2(0,2\pi)$ and,
therefore, \be
c_l=\frac{1}{2\pi}\int_0^{2\pi}\,C(p)\,e^{-ipl}\,dp=\frac{1}{2\pi}\int_0^{2\pi}\,
\frac{e^{-ipl}\,dp}{\sqrt{\alpha(p)}}\label{II10}\en with a
particular choice of phase for $C(p)$. Now, due to the regularity of
the function $1/\sqrt{\alpha(p)}$ and to its $2\pi$-periodicity, it
is a standard exercise in Fourier series theory to check that $c_l$
goes to zero when $l$ diverges faster than any inverse power of $l$.
Therefore the series in (\ref{II3}), (\ref{II5}) and (\ref{II9}) all
converge, and we conclude that the set
$\M_1=\{\varphi_n,\,n\in\Bbb{Z}\}$ is an orthonormal set in
$\Hil_1$. A natural question is now the following: is $\M_1$
complete in $\Hil_1$? To answer this question we give here the
following proposition, which gives a necessary and sufficient
condition for $\M_1$ to be complete in $\Hil_1$. In the proof of
this proposition we will use the fact that, under the assumptions of
the statement, $X$ is self adjoint and maps $\Hil_1$ into itself.
The proof of this claim is a simple exercise and is left to the
reader.

\begin{prop}
Suppose that $\{a_j\}\in l^1(\Bbb{Z})$ and  that $\alpha(p)\neq 0$
for all $p\in[0,2\pi[$. Then $\M_1$ is complete in $\Hil_1$ if and
only if $X$ admits a bounded inverse.
\end{prop}
\begin{proof}
Let $h\in\Hil_1$ be orthogonal to all the $\varphi_n$'s,
$n\in\Bbb{Z}$. Then, because of (\ref{II4}), we have
$0=<h,\varphi_n>=<h,Xf_n>=<Xh,f_n>$ for all $n\in\Bbb{Z}$. But
$\Nc_1$ is complete in $\Hil_1$ and $Xh\in\Hil_1$ since $h\in\Hil_1$
and $X:\Hil_1\rightarrow\Hil_1$. Therefore $Xh=0$. Since $X$ is
invertible, then $h=0$ and, as a consequence, $\M_1$ is complete.

\vspace{1mm}

Let us prove the converse statement. Since $\M_1$ is complete in
$\Hil_1$ and since $f_0\in\Hil_1$ then we can write \be
f_0=\sum_{l\in\Bbb{Z}}\,d_l\,\varphi_l,\label{II11}\en and $\{d_l\}$
satisfies the sum rule $\sum_{l\in\Bbb{Z}}\,|d_l|^2=1$ because
$\M_1$ is an o.n. complete set and $f_0$ is normalized. Moreover we
have
$a_j=<A^jf_0,f_0>=<f_j,f_0>=\sum_{l,k\in\Bbb{Z}}\,\overline{d_l}\,d_k\,<\varphi_{l+j},\varphi_k>=
\sum_{l\in\Bbb{Z}}\,\overline{d_l}\,d_{l+j}$ which, introducing the
function $D(p)=\sum_{l\in\Bbb{Z}}\,d_l\,e^{ipl}\in\Lc^2(0,2\pi)$,
 becomes $|D(p)|^2=\alpha(p)$ a.e. in $[0,2\pi[$. Therefore we
get \be
d_n=\frac{1}{2\pi}\int_0^{2\pi}\,D(p)\,e^{-ipn}\,dp=\frac{1}{2\pi}\int_0^{2\pi}\,\sqrt{\alpha(p)}
\,e^{-ipn}\,dp\label{II12}\en with a particular choice of phase for
$D(p)$. Because of our assumption on $a_j$ it follows that the
series for $\alpha(p)$ converges uniformly and define a positive
$C^\infty$  function which is also $2\pi-$periodic. These features
are also shared by $\sqrt{\alpha(p)}$ and therefore $d_n$ decreases
to zero faster than any inverse power of $n$, as
$n\rightarrow\infty$.

Now, since
$f_n=A^nf_0=A^n\left(\sum_{l\in\Bbb{Z}}\,d_l\,\varphi_l\right)=
\left(\sum_{l\in\Bbb{Z}}\,d_l\,A^l\right)\varphi_n$, and since
$\sum_{l\in\Bbb{Z}}\,d_l\,A^l$ surely converges uniformly, it is
clear that this defines a new bounded operator which is exactly the
inverse of $X$, namely $X^{-1}=\sum_{l\in\Bbb{Z}}\,d_l\,A^l$.

\end{proof}

\vspace{2mm}

{\bf Remark:} the requirement $\alpha(p)\neq 0$ for all
$p\in[0,2\pi[$ is used above to ensure that the operator $X$
exists and is bounded, as it can be deduced from the asymptotic
behavior of the coefficients $c_l$'s.

\vspace{1mm}

An interesting result relating the coefficients of the two
expansions in (\ref{II3}) and (\ref{II11}), which may be considered
as the inverse one of the other, is given by the following sum rule:
\be \sum_{n\in{\Bbb{Z}}}\overline{c_n}\,d_n=1\label{II13}\en The
proof makes use of the Poisson summation rule, \cite{dau},
$\sum_{n\in{\Bbb{Z}}}e^{ixan}=\frac{2\pi}{|a|}\sum_{n\in{\Bbb{Z}}}\delta\left(x-\frac{2\pi}{a}n\right)$,
$a\neq 0$, and goes as follows:
$$
\sum_{n\in{\Bbb{Z}}}\overline{c_n}\,d_n=\sum_{n\in{\Bbb{Z}}}\left(\frac{1}{2\pi}\int_0^{2\pi}\,
\frac{e^{ipn}}{\sqrt{\alpha(p)}}
\,dp\right)\,\left(\frac{1}{2\pi}\int_0^{2\pi}\,\sqrt{\alpha(q)}
\,e^{-iqn}\,dq\right)=
$$
$$
=\frac{1}{(2\pi)^2}\int_0^{2\pi}\,dp\int_0^{2\pi}\,dq\sqrt{\frac{\alpha(q)}{\alpha(p)}}\,
\sum_{n\in{\Bbb{Z}}}e^{in(p-q)}=$$
$$=\frac{1}{2\pi}\int_0^{2\pi}\,dp\int_0^{2\pi}\,dq\sqrt{\frac{\alpha(q)}{\alpha(p)}}\,
\sum_{n\in{\Bbb{Z}}}\delta(p-q-2\pi
n)=\frac{1}{2\pi}\int_0^{2\pi}\,dp\int_0^{2\pi}\,dq\sqrt{\frac{\alpha(q)}{\alpha(p)}}\,
\delta(p-q)=1,
$$
because the only effective contribution  arising here from
$\sum_{n\in{\Bbb{Z}}}\delta(p-q-2\pi n)$ comes from $n=0$, since
$p,q\in[0,2\pi[$.

\subsection{Preliminary examples}

Let $f_0(x)=\chi_{[0,a[}(x)$ be the characteristic function in the
interval $[0,a[$, with $a>0$, and let $A$ be the following
translation operator: $A=e^{-i\hat p}$. We have
$$\Nc_1=\{f_{n}(x):=A^{n}f_0(x)=\chi_{[n,n+a[}(x),\,\,
n\in\Bbb{Z} \}.$$ We want to see what our procedure produces
starting with this set. For that, it is convenient to consider
separately the cases $a<1$, $a=1$ and $a>1$. Let us start with the
easiest case, $a=1$. In this case the set $\Nc_1$ is already made of
o.n.  functions, and therefore we expect that the set $\M_1$
coincides with $\Nc_1$. Indeed this is what happens, since
$a_j=<f_j,f_0>=\delta_{j,0}$. Therefore $\alpha(p)=1$, which is
obviously never zero, and $c_l=\delta_{l,0}$, see (\ref{II10}). From
(\ref{II4}) we deduce that $\varphi_n(x)=f_n(x)$ for all integer
$n$. It is clear that both $X$ and $X^{-1}$ exist, and they are both
equal to the identity operator.

Just a little less trivial is the situation when $a<1$. In this
case, in fact, the set $\Nc_1$ is still made of orthogonal
functions, since each $f_{n}(x)=\chi_{[n,n+a[}(x)$ does not
overlap with any other $f_{k}(x)=\chi_{[k,k+a[}(x)$, if $k\neq n$.
However none of these functions is normalized so that we may
expect that our procedure simply {\em cures} this feature. Indeed
we have $a_j=<f_j,f_0>=a\delta_{j,0}$, so that $\alpha(p)=a$,
which is again never zero, and
$c_l=\frac{1}{\sqrt{a}}\,\delta_{l,0}$. Therefore
$\varphi_n(x)=\frac{1}{\sqrt{a}}\,f_n(x)$ for all integer $n$. Of
course these are now orthogonal functions with norm equal to 1. It
is finally clear that both $X$ and $X^{-1}$ exist, and we find
$X=\frac{1}{\sqrt{a}}\,\Id$ and $X^{-1}=\sqrt{a}\,\Id$.

Surely more interesting is the case $a>1$. We restrict ourselves,
for the time being, to $1<a<2$. The overlap coefficients $a_j$ can
  be written as
$a_j=a\,\delta_{j,0}+(a-1)\left(\delta_{j,-1}+\delta_{j,1}\right)$,
so that $\alpha(p)=a+2(a-1)\cos(p)$. This is a nonnegative, real and
$2\pi$-periodic function, as expected, and furthermore it is never
zero in $[0,2\pi[$ since it has a minimum in $p=\pi$ and
$\alpha(\pi)=2-a>0$. If we fix, just to be concrete,
$a=\frac{3}{2}$, we can compute analytically
$\sum_{l\in{\Bbb{Z}}}\,|c_l|^2=\frac{1}{2\pi}\int_0^{2\pi}\frac{dp}{\alpha(p)}=\frac{2}{\sqrt{5}}$.
Therefore the sequence $\{c_l\}$ belongs to $l^2({\Bbb{Z}})$, as it
was to be expected because of the absence of zeroes of $\alpha(p)$.
As a matter of fact, it is quite easy to check also numerically that
both $c_l$ and $d_l$ decrease very fast for increasing $l$: already
for $|l|\geq 5$ we find $|c_l|\simeq 10^{-3}$ and $|d_l|\simeq
2\cdot10^{-4}$. It is also easy to check that the sum rule in
(\ref{II13}) is satisfied. This same analysis can be extended to
$a\geq 2$. One can check that there are values of the parameter $a$
for which, e.g., $\{c_l\}$ belongs to $l^2({\Bbb{Z}})$, and other
values of $a$, for which $\{c_l\}\notin l^2({\Bbb{Z}})$.

For instance, if $a=2$ the overlap coefficients are the same as
for $a\in]1,2[$,
$a_j=a\,\delta_{j,0}+(a-1)\left(\delta_{j,-1}+\delta_{j,1}\right)=2
\,\delta_{j,0}+\left(\delta_{j,-1}+\delta_{j,1}\right)$, so that
$\alpha(p)=2+2\cos(p)$. This  is zero for $p=\pi$ and one can
check that $\sum_{l\in{\Bbb{Z}}}\,|c_l|^2=+\infty.$ So the same
example produces different behavior depending on the value of $a$.
We will recover this same feature in the next section, in the
construction of the so-called {\em orhogonal coherent-states}.

\vspace{2mm}

Another interesting and easy example is the following: let
$f_0(x)=\chi_{[0,1[}(x)$ and let $A$ be the following dilatation
operator: $(Ah)(x)=\sqrt{2}\,h(2x)$,
$\forall\,h(x)\in\Lc^2(\Bbb{R})$. Then the set $\Nc_1$ turns out to
be
$$
\Nc_1=\left\{f_n(x)=2^{n/2}f(2^nx)=2^{n/2}\left\{
\begin{array}{ll}
1, \quad\mbox{if }0< x\leq 2^{-n},  \\
0, \quad\mbox{otherwise},
\end{array}
\right.\quad n\in\Bbb{Z}\right\}
$$
In this case all the overlap coefficients $a_j$ are different from
zero. Indeed we get $a_j=2^{-|j|/2}$, for all $j\in\Bbb{Z}$. Since
$\left|\frac{e^{\pm ip}}{\sqrt{2}}\right|=\frac{1}{\sqrt{2}}<1$,
it is easy to compute the analytic expression of $\alpha(p)$ and
it turns out that $\alpha(p)=\frac{1}{3-2^{3/2}\cos(p)}$. The
minimum of $\alpha(p)$ is found again for $p=\pi$, and
$\alpha(\pi)=\frac{1}{3+2^{3/2}}\simeq 0.1716$, which is different
from zero. Moreover we find that
$\max(\alpha(p))=\alpha(0)=\frac{1}{3-2^{3/2}}\simeq 5.8284$. The
$\|.\|_2$-norm of the sequence $\{c_l\}$ can be computed
analytically and we find
$\sum_{l\in{\Bbb{Z}}}\,|c_l|^2=\frac{1}{2\pi}\int_0^{2\pi}\frac{dp}{\alpha(p)}=3$.
Again, it is quite easy to find numerically the value of the
coefficients $c_l$ and $d_l$, to check that they both converge to
zero quite fast, and that (\ref{II13}) is satisfied. Further, one
can use these coefficients to define the {\em new} o.n. vectors
using (\ref{II3}) and (\ref{II4}).

\section{Coherent states}

This section is devoted to a  more interesting example involving
coherent states, \cite{aag}. We will see that the set of coherent
states fits the general discussion of Section II, and we will show
how and when the orthonormalization procedure works.

Let $\hat q$ and $\hat p$ be the position and momentum operators
on a Hilbert space $\Hil$, $[\hat q,\hat p]=i\Id$, and let us now
introduce the following unitary operators: \be U(\underline
n)=e^{ia(n_1\hat q-n_2\hat p)}, \quad D(\underline
n)=e^{z_{\underline{n}} b^\dagger-\overline{z}_{\underline n}
b},\quad T_1:=e^{ia\hat q},\quad T_2:=e^{-ia\hat p}.\label{31}\en
Here $a$ is a real constant satisfying $a^2=2\pi L$ for some
$L\in\Bbb{N}$, while $z_{\underline{n}}$ and $b$ are related to
${\underline{n}}=(n_1,n_2)$ and $\hat q$, $\hat p$ via the
following equalities: \be
z_{\underline{n}}=\frac{a}{\sqrt{2}}(n_2+in_1),\quad
b=\frac{1}{\sqrt{2}}(\hat q+i\hat p). \label{32}\en With these
definitions it is clear that \be U(\underline n)=D(\underline
n)=(-1)^{Ln_1n_2}T_1^{n_1}T_2^{n_2}=(-1)^{Ln_1n_2}T_2^{n_2}T_1^{n_1},\label{33}\en
where we have also used the commutation rule $[T_1,T_2]=0$.

Let $\varphi_{\underline{0}}$ be the vacuum of $b$,
$b\varphi_{\underline{0}}=0$, and let us define the following {\em
coherent states}: \be
\varphi_{\underline{n}}:=T_1^{n_1}T_2^{n_2}\varphi_{\underline{0}}=T_2^{n_2}T_1^{n_1}
\varphi_{\underline{0}} =(-1)^{Ln_1n_2}U(\underline
n)\varphi_{\underline{0}}=(-1)^{Ln_1n_2}D(\underline
n)\varphi_{\underline{0}}.\label{34}\en It is  very well known that
the set of these vectors,
$\C=\{\varphi_{\underline{n}},\,\underline{n}\in{\Bbb{Z}}^2\}$,
satisfies, among the others, the following properties:
\begin{enumerate} \item $\C$ is invariant under the action of
$T_j^{n_j}$, $j=1,2$;
\item each $\varphi_{\underline{n}}$ is an eigenstate of $b$:
$b\varphi_{\underline{n}}=z_{\underline{n}}\,\varphi_{\underline{n}}$;\item
they satisfy the {\em resolution of the identity}
$\sum_{\underline{n}\in{\Bbb{Z}}^2}\,|\varphi_{\underline{n}}><\varphi_{\underline{n}}|=\Id$;\item
They saturate the Heisenberg uncertainty principle: let $(\Delta
X)^2=<X^2>-<X>^2$ for $X=\hat q,\hat p$, then $\Delta \hat q\,\Delta
\hat p=\frac{1}{2}$.
\end{enumerate}
However, it is also well known  that they are not mutually
orthogonal. Indeed we have: \be
I_{\underline{n}}:=<\varphi_{\underline{n}},\varphi_{\underline{0}}>=(-1)^{Ln_1n_2}\,e^{-\frac{\pi}{2}L(n_1^2
+n_2^2)}.\label{35}\en Of course, for large $L$  the set $\C$ can be
considered as {\em approximately orthogonal}, since
$I_{\underline{n}}\simeq 0$ for all $\underline{n}\neq \underline
0$. On the contrary, for small $L$,  the overlap between neirest
neighboring vectors is significantly different from zero.

Our aim is to construct a family of vectors $\E$ which shares with
$\C$ most of the above features and which, moreover, is made of
orthonormal vectors. We will show that this is possible, in suitable
Hilbert spaces, if $L>1$, while the procedure discussed in Section
II fails for $L=1$.

We start our analysis with some consideration concerning the set
$\C$. For this we will make use of the results on the generalized
$(k,q)$ representation presented in the Appendix. Since most of
our results will depend on the value of $L$, i.e. on the value of
$a^2$, from now on we replace $\varphi_{\underline{n}}$ with
$\varphi_{\underline{n}}^{(L)}$, and $\C$ with $\C^{(L)}$.
However, it is important to stress that, due to its definition,
$\varphi_{\underline{0}}$ does not depend on $L$, while all the
 vectors $\varphi_{\underline{n}}^{(L)}=T_1^{n_1}T_2^{n_2}\varphi_{\underline{0}}$ do.
  Our first result is the following:

\begin{prop} The set $\C^{(L)}$ is complete in $\Hil$ if and only if
$L=1$.\end{prop}
\begin{proof}
The proof of this statement extends the analogous proof given in
\cite{bgz}: let $h\in\Hil$ be a vector orthogonal to
$\varphi_{\underline{n}}^{(L)}$ for all $\underline n
\in{\Bbb{Z}}^2$: $<h,\varphi_{\underline{n}}^{(L)}>=0$
$\forall\,\underline n \in{\Bbb{Z}}^2$. Using the functions
$\Phi_{k,q}^{(A,a)}(x)$ introduced in (\ref{28}), $A>0$ fixed and
$(k,q)\in\Box^{(A)}:=\left[0,\frac{2\pi}{A}\right[\times[0,a[$, and
their properties, we deduce that
$$
0=<h,\varphi_{\underline{n}}^{(L)}>=\int\int_{\Box^{(A)}}\,<h,\Phi_{k,q}^{(A,a)}
><\Phi_{k,q}^{(A,a)},\varphi_{\underline{n}}^{(L)}>\,dk\,dq,
$$
see (\ref{211}), and since
$$<\Phi_{k,q}^{(A,a)},\varphi_{\underline{n}}^{(L)}>=<\Phi_{k,q}^{(A,a)},T_1^{n_1}T_2^{n_2}
\varphi_{\underline{0}}>= e^{-iqan_1+ikAn_2}<\Phi_{k,q}^{(A,a)},
\varphi_{\underline{0}}>,$$ see (\ref{29}), we find
$$
0=\int_0^{2\pi/A}\,dk\,e^{ikAn_2}\,\int_0^{a}\,dq\,e^{-iqan_1}\,C(k,q),
$$
for all $\underline{n}\in {\Bbb{Z}}^2$. Here we have introduced
the function $C(k,q):=<h,\Phi_{k,q}^{(A,a)}
><\Phi_{k,q}^{(A,a)},\varphi_{\underline{0}}>$. In this way the
problem of the completeness of the set $\C^{(L)}$ has been
replaced by the problem of completeness of the set
$\D^{(L)}:=\left\{e^{-iqan_1+ikAn_2},\,\underline{n}\in{\Bbb{Z}}^2\right\}$
in $\Lc^2(\Box^{(A)})$. It is now easy to prove that, if $L>1$,
the function $s(k,q)=e^{iqa/L}$ belongs to $\Lc^2(\Box^{(A)})$, is
different from zero a.e. in $\Box^{(A)}$, and it is orthogonal to
all the functions in $\D^{(L)}$. Therefore, if $L>1$, $\D^{(L)}$
is not complete and as a consequence, $\C^{(L)}$ is not complete
either.

If $L=1$ the completeness of $\D^{(1)}$ is a well known fact in
the theory of the Fourier series. Moreover, since
$<\Phi_{k,q}^{(A,a)},\varphi_{\underline{0}}>\neq 0$ a.e. in
$\Box^{(A)}$, \cite{bgz}, we conclude that $h=0$: $\C^{(1)}$ is
complete in $\Hil$.
\end{proof}

\vspace{2mm}

Let us now define, for each $L\geq 1$, the following set: \be
h_L:=\mbox{linear
span}\overline{\left\{\varphi_{\underline{n}}^{(L)},\,\underline{n}\in{\Bbb{Z}}^{
2}\right\}}^{\|.\|}. \label{36}\en It is clear that $h_1=\Hil$,
while, for $L>1$, $h_L\subset \Hil$. It is further clear that
$h_L$ is an Hilbert space for each $L$, since it is a closed
subspace of $\Hil$. Furthermore, we can associate to $h_L$ two
different Hilbert spaces of functions, obtained by {\em
projecting} the vectors of $h_L$ in the coordinate  or in the
$(k,q)$-representation. We have, see also the appendix,
$$
l^2_L({\Bbb{R}}):=\left\{f(x)\in\Lc^2({\Bbb{R}}):\,\exists f\in h_L:
\, f(x)=<\xi_x,f> \right\},
$$
and
$$
l^2_L(\Box^{(A)}):=\left\{f(k,q)\in\Lc^2(\Box^{(A)}):\,\exists f\in
h_L: \, f(k,q)=<\Phi_{k,q}^{(A,a)},f> \right\}.
$$
From what we have discussed above, it is clear that
$l^2_L({\Bbb{R}})$ and $l^2_L(\Box^{(A)})$ are closed subsets of
$\Lc^2_L({\Bbb{R}})$ and $\Lc^2_L(\Box^{(A)})$ respectively, so that
they are Hilbert spaces, too.

The problem we want to discuss here is the following: {\em is it
possible to produce, starting from $\C^{(L)}$, a set of vectors
which are still coherent (at a certain extent) and which are
mutually orthogonal }? It is clear that this last requirement is not
compatible with what one usually calls {\em coherent states},
\cite{klauder}. However we will see that adopting here the procedure
discussed in Section II a rather non-trivial structure emerges.

We start extending formula (\ref{II4}) to the present settings: \be
\Psi_{\underline{n}}^{(L)}:=\sum_{\underline{k}\in{\Bbb{Z}}^2}\,c_{\underline{k}}^{(L)}\,
\varphi_{\underline{k}+\underline{n}}^{(L)} \label{37} \en Of course
this means that
$\Psi_{\underline{0}}^{(L)}:=\sum_{\underline{k}\in{\Bbb{Z}}^2}\,c_{\underline{k}}^{(L)}\,
\varphi_{\underline{k}}^{(L)}$ and, because of the commutativity of
$T_1$ and $T_2$, that \be
\Psi_{\underline{n}}^{(L)}=T_1^{n_1}\,T_2^{n_2}\Psi_{\underline{0}}^{(L)}.
\label{38}\en Therefore the new set constructed in this way,
$\E^{(L)}:=\{\Psi_{\underline{n}}^{(L)},\,\underline{n}\in{\Bbb{Z}}^2\}$,
is invariant under the action of $T_1$ and $T_2$, exactly as the set
$\C^{(L)}$, independently of the choice of the coefficients of the
expansion $c_{\underline{k}}^{(L)}$. Useless to say, in order to
have a converging expansion in (\ref{37}), the following inequality
must be satisfied: \be
\sum_{\underline{k},\underline{s}\in{\Bbb{Z}}^2}\,(-1)^{L(k_1-s_1)(k_2-s_2)}\,e^{-\frac{\pi}{2}\,L((k_1-s_1)^2
+(k_2-s_2)^2)}\,\overline{c_{\underline{k}}^{(L)}}\,c_{\underline{s}}^{(L)}<\infty
\label{38bis}\en which is equivalent to  require that
$\|\Psi_{\underline{n}}^{(L)}\|=\|\Psi_{\underline{0}}^{(L)}\|<\infty$.
It is clear then, because of the Schwarz inequality, that if
(\ref{38bis}) holds then all the scalar products
$<\Psi_{\underline{n}}^{(L)},\Psi_{\underline{s}}^{(L)}>$ are well
defined. Of course the coefficients $c_{\underline{s}}^{(L)}$ must
not be chosen freely: they are fixed by requiring that the vectors
in the set $\E^{(L)}$ are  orthonormal:
$<\Psi_{\underline{n}}^{(L)},\Psi_{\underline{s}}^{(L)}>=\delta_{\underline{n},
\underline{s} }$. This will fix (not uniquely!) the value of the
$c_{\underline{s}}^{(L)}$'s, with a procedure which extends what we
have discussed in the previous section and which is also close to
the one used in \cite{bms} in a different context. We will also
check that the set $\E^{(L)}$ is complete in $h_L$.

\vspace{2mm}

In order to deduce the expression for $c_{\underline{s}}^{(L)}$ we
start observing that in order to have orthogonality among all the
$\Psi_{\underline{n}}^{(L)}$, it is enough to require that
$<\Psi_{\underline{n}}^{(L)},\Psi_{\underline{0}}^{(L)}>=\delta_{\underline{n},
\underline{0} }$ for all $\underline{n}\in{\Bbb{Z}}^2$. Indeed, if
this is satisfied, then the invariance under translations of the
set $\E^{(L)}$ implies also that
$<\Psi_{\underline{n}}^{(L)},\Psi_{\underline{s}}^{(L)}>=\delta_{\underline{n},
\underline{s} }$ for all $\underline{n}, \underline{s}
\in{\Bbb{Z}}^2$. Using expansion (\ref{37}) we find that
\be<\Psi_{\underline{n}}^{(L)},\Psi_{\underline{0}}^{(L)}>=\sum_{\underline{k},\underline{s}\in{\Bbb{Z}}^2}
\,\overline{c_{\underline{l}}^{(L)}}\,c_{\underline{s}}^{(L)}\,I_{\underline{l}+\underline{n}-\underline{s}}=
\delta_{\underline{n}, \underline{0} },\label{311}\en which is
equivalent to the following equation: \be
F_L(\underline{P})\,|C_L(\underline{P})|^2=1,\quad \mbox{a.e. in}
[0,2\pi[\times[0,2\pi[,\label{312}\en where \be
F_L(\underline{P}):=\sum_{\underline{k}\in{\Bbb{Z}}^2}\,
I_{\underline{k}}\,e^{i\underline{P}\cdot\underline{k}} \quad
\mbox{and}\quad
C_L(\underline{P}):=\sum_{\underline{k}\in{\Bbb{Z}}^2}\,
c_{\underline{k}}^{(L)}\,e^{i\underline{P}\cdot\underline{k}}
\label{313}\en It is clear now that the coefficients can be
recovered via the formula \be
c_{\underline{k}}^{(L)}=\frac{1}{(2\pi)^2}\,\int_0^{2\pi}\int_0^{2\pi}\,\frac{e^{-i\underline{P}
\cdot\underline{k}}}{\sqrt{F_L(\underline{P})}}\,d\underline{P},\label{314}\en
which corresponds to a special choice of the phase of the function
$C_L(\underline{P})$. We will show in a moment that this integral
does not need to exist in general and, even if it exists, there is
no reason a priori to ensure that the coefficients
$c_{\underline{k}}^{(L)}$'s satisfy condition (\ref{38bis}). This
is a consequence of the non orthogonality of the set $\C^{(L)}$
and of the procedure we are adopting. However, under simple
conditions, it is possible to analyze the asymptotic behavior of
the $c_{\underline{k}}^{(L)}$'s for $\underline{k}$ diverging
using more or less standard techniques which relates this behavior
to the analytic features of $F_L(\underline{P})$. First we see
that, since \be
F_L(\underline{P})=\sum_{\underline{m}\in{\Bbb{Z}}^2}\,(-1)^{L\,m_1\,m_2}\,e^{-\frac{\pi}{2}\,L(m_1^2
+m_2^2)}\,e^{i\underline{P}\cdot\underline{m}}, \label{315}\en
$F_L$ can be rewritten in terms of the Jacobi $\theta_3$ function
as follows: \be
F_L(\underline{P})=\theta_3\left(\frac{P_1}{2},e^{-\frac{\pi}{2}L}\right)
\theta_3\left(P_2,e^{-2\pi
L}\right)+e^{iP_2-\frac{\pi}{2}L}\,\theta_3\left(\frac{P_1+\pi L
}{2},e^{-\frac{\pi}{2}L}\right) \theta_3\left(P_2+i\pi L,e^{-2\pi
L}\right). \label{316}\en

We have also found a different expression for
$F_L(\underline{P})$, again in terms of $\theta_3$, which we
report here just for completeness: \be F_L(\underline{P})=e^{i\pi
LD }\theta_3\left(\frac{P_1}{2},e^{-\frac{\pi}{2}L}\right)
\theta_3\left(\frac{P_2}{2},e^{-\frac{\pi}{2}L}\right),
\label{317}\en
 where $D$ is the differential operator defined as
$D=\left(-i\frac{\partial}{\partial
P_1}\right)\left(-i\frac{\partial}{\partial P_2}\right)$. A nice
feature of formula (\ref{317}), when compared to (\ref{316}), is
that (\ref{317}) is manifestly invariant under the exchange
$P_1\leftrightarrow P_2$, as the original expression in (\ref{315}),
while the other is not.

The function $F_L(P_1,P_2)$ is surely nonnegative, since it has to
satisfy (\ref{312}), and $2\pi$-periodic:
$F_L(P_1+2\pi,P_2+2\pi)=F_L(P_1,P_2)$ a.e. It is also infinitely
differentiable, for all $L\geq 1$. However, since $F_1(\pi,\pi)=0$,
there is no reason a priori for the integral in (\ref{314})  to be
convergent if $L=1$ and, even if this happens, there is no reason
for the related $\{c_{\underline{k}}^{(L)}\}$ to satisfy condition
(\ref{38bis}). For this reason it is more convenient to consider
separately the two situations $L=1$ and $L>1$.

\subsection{What if $L>1$?}

If $L>1$ it is possible to prove that the function
$F_L(\underline{P})$ has no zero at all. Indeed, if we write
$F_L(\underline{P})=1+F_L^o(\underline{P})$,
$F_L^o(\underline{P})=\sum_{\underline{m}\in{\Bbb{Z}}^2\setminus{(0,0)}}\,(-1)^{L\,m_1\,m_2}\,e^{-\frac{\pi}{2}\,L(m_1^2
+m_2^2)}\,e^{i\underline{P}\cdot\underline{m}}$, we deduce that
$$
|F_L^o(\underline{P})|\leq\sum_{\underline{m}\in{\Bbb{Z}}^2\setminus{(0,0)}}\,e^{-\frac{\pi}{2}\,L(m_1^2
+m_2^2)}=\sum_{\underline{m}\in{\Bbb{Z}}^2}\,e^{-\frac{\pi}{2}\,L(m_1^2
+m_2^2)}-1=\left(\theta_3\left(0,e^{-\frac{\pi}{2}L}\right)\right)^2-1,
$$
for all $\underline{P}\in[0,2\pi[\times[0,2\pi[$. The right-hand
side can be easily computed for different values of $L$. We get:
$|F_1^o(\underline{P})|\leq 1.01497$, while
$|F_2^o(\underline{P})|\leq 0.180341$, $|F_3^o(\underline{P})|\leq
0.036256$ and so on.  As we can see, $F_L(\underline{P})$ can only
be zero for some $\underline{P}$ if $L=1$, and this is exactly what
happens for $\underline P=(\pi,\pi)$, while for $L\geq 2$
$F_L(\underline{P})$ is strictly positive.

With this in mind we conclude that for $L>1$ the function
$C_L(\underline{P})=\frac{1}{\sqrt{F_L(\underline{P})}}$ is always
well defined, belongs to $C^\infty$, and is $(2\pi,2\pi)$-periodic
together with all its derivatives. A standard argument allows us to
conclude therefore that the coefficients $c_{\underline{k}}^{(L)}$
in (\ref{314}) go to zero faster than any inverse power of
$\|\underline{k}\|=\sqrt{k_1^2+k_2^2}$. Let us now put, for
$N\in\Bbb{N}$,$\Psi_{\underline{0},\,N}^{(L)}=\sum_{\|\underline{k}\|\leq
N}\,c_{\underline{k}}^{(L)}\,\varphi_{\underline{k}}^{(L)}$, and let
$N>M$. Then we have
$$
\|\Psi_{\underline{0},\,N}^{(L)}-\Psi_{\underline{0},\,M}^{(L)}\|\leq
\sum_{M<\|\underline{k}\|\leq
N}\,|c_{\underline{k}}^{(L)}|\,\|\varphi_{\underline{k}}^{(L)}\|=\sum_{M<\|\underline{k}\|\leq
N}\,|c_{\underline{k}}^{(L)}|\rightarrow 0
$$
when $M,N\rightarrow\infty$, due to the asymptotic behavior of
$c_{\underline{k}}^{(L)}$. Since $h_L$ is complete, the sequence
$\{\Psi_{\underline{0},\,N}^{(L)}\}$ is convergent to an element
of $h_L$, which is clearly $\Psi_{\underline{0}}^{(L)}$. The same
argument can be repeated to check that
$\Psi_{\underline{n}}^{(L)}$ is well defined and belongs to $h_L$.
Alternatively, we can simply observe that since
$\Psi_{\underline{0}}^{(L)}$ belongs to $h_L$, and since $h_L$ is
invariant under the action of $T_1$ and $T_2$, also
$\Psi_{\underline{n}}^{(L)}=T_1^{n_1}T_2^{n_2}\Psi_{\underline{0}}^{(L)}$
belongs to $h_L$.

Going back to (\ref{37}), if we introduce  an operator $X_L$ as in
(\ref{II5}), \be
X_L=\sum_{\underline{k}\in{\Bbb{Z}}^2}\,c_{\underline{k}}^{(L)}\,T_1^{k_1}\,T_2^{k_2},
\label{318}\en this can be rewritten as \be
\Psi_{\underline{k}}^{(L)}=X_L\,\varphi_{\underline{k}}^{(L)}\label{319}\en
for all $\underline{k}\in{\Bbb{Z}}^2$. This is exactly the
analogous of equation (\ref{II4}). The operator $X_L$ is, for
$L>1$, bounded and self-adjoint. Indeed we have
$$
\|X_L\|\leq
\sum_{\underline{k}\in{\Bbb{Z}}^2}\,|c_{\underline{k}}^{(L)}|\,\|T_1^{k_1}\|\,\|T_2^{k_2}\|=
\sum_{\underline{k}\in{\Bbb{Z}}^2}\,|c_{\underline{k}}^{(L)}|<\infty
$$
again because of the asymptotic behavior of
$c_{\underline{k}}^{(L)}$. Moreover we have, since formula
(\ref{314}) implies that
$\overline{c_{\underline{k}}^{(L)}}=c_{-\underline{k}}^{(L)}$,
$$
X_L^\dagger=\sum_{\underline{k}\in{\Bbb{Z}}^2}\,\overline{c_{\underline{k}}^{(L)}}
\,{T_1^{k_1}}^\dagger\,{T_2^{k_2}}^\dagger=\sum_{\underline{k}\in{\Bbb{Z}}^2}\,c_{-\underline{k}}^{(L)}
\,T_1^{-k_1}\,T_2^{-k_2}=\sum_{\underline{n}\in{\Bbb{Z}}^2}\,c_{\underline{n}}^{(L)}
\,T_1^{n_1}\,T_2^{n_2}=X_L.
$$

We will show in the last part of this subsection that $X_L$ admits a
bounded inverse, as soon as $L>1$. At this stage we simply assume
that this is so: $X_L^{-1}$ exists and belongs to $B(h_L)$, the set
of all the bounded operators on $h_L$. This assumption allows us to
prove that the set $\E^{(L)}$ is complete in $h_L$, just extending
the same argument of the previous section. Indeed, let $g\in h_L$ be
such that $<g,\Psi_{\underline{n}}^{(L)}>=0$ for all
$\underline{n}\in{\Bbb{Z}}^2$. Then we have,
$\forall\,\underline{n}\in{\Bbb{Z}}^2$,
$0=<g,X_L\varphi_{\underline{n}}^{(L)}>=<X_L\,g,\varphi_{\underline{n}}^{(L)}>$.
Since the set $\C^{(L)}$ is complete in $h_L$ by construction, then
we must have $X_L\,g=0$ or, applying $X_L^{-1}$, $g=0$.

\vspace{2mm}

{\bf Remark:} of course it is necessary to check that $X_L g\in
h_L$ for any $g\in h_L$, but this is a simple exercise and is left
to the reader. It is also easy to reverse this statement and to
check that, under additional conditions that remind those of
Proposition 1, if $\E^{(L)}$ is complete in $h_L$ then the
operator $X_L$ must admit a bounded inverse.

\vspace{1mm}

Once we have proven that the set $\E^{(L)}$ is complete in $h_L$ we
can expand each vector $\varphi_{\underline{n}}^{(L)}$ in terms of
the $\Psi_{\underline{n}}^{(L)}$ in a translationally invariant way:
\be
\varphi_{\underline{n}}^{(L)}=\sum_{\underline{k}\in{\Bbb{Z}}^2}\,\alpha_{\underline{k}}^{(L)}\,
\Psi_{\underline{n}+\underline{k}}^{(L)}\label{320}\en As we have
already seen in Section II, the analysis of these coefficients is,
in a sense, much simpler than that of the $c_{\underline{k}}^{(L)}$,
since we can here use the Parseval equality because of the
orthonormality of the set $\E^{(L)}$. For instance we have
$1=\|\varphi_{\underline{n}}^{(L)}\|^2=
\sum_{\underline{k}\in{\Bbb{Z}}^2}\,|\alpha_{\underline{k}}^{(L)}|^2$,
which proves that $\{\alpha_{\underline{k}}^{(L)}\}$ belongs to
$l^2({\Bbb{Z}}^2)$ for all $L>1$. Moreover, using (\ref{320}) and
(\ref{35}) (and replacing $I_{\underline{n}}$ with
$I_{\underline{n}}^{(L)}$), we find that
$I_{\underline{n}}^{(L)}=<\varphi_{\underline{n}}^{(L)},
\varphi_{\underline{0}}^{(L)}>
=\sum_{\underline{k},\,\underline{s}\in{\Bbb{Z}}^2}\,\overline{\alpha_{\underline{k}}^{(L)}}\,
\alpha_{\underline{s}}^{(L)}
<\Psi_{\underline{n}+\underline{k}}^{(L)},
\Psi_{\underline{s}}^{(L)}>=
\sum_{\underline{k}\,\in{\Bbb{Z}}^2}\,\overline{\alpha_{\underline{k}}^{(L)}}\,
\alpha_{\underline{k}+\underline{n}}^{(L)} $. If we now multiply
both sides of this equality for
$e^{i\underline{P}\cdot\underline{n}}$ and sum up on
$\underline{n}\in {\Bbb{Z}}^2$, we get \be
F_L(\underline{P})=|G_L(\underline{P})|^2,\quad \mbox{a.e. in}\,
[0,2\pi[\times[0,2\pi[,\label{321}\en where $F_L(\underline{P})$ has
been defined in (\ref{313}), while \be
G_L(\underline{P})=\sum_{\underline{k}\,\in{\Bbb{Z}}^2}\,\alpha_{\underline{k}}^{(L)}\,
e^{i\underline{P}\cdot\underline{k}}.\label{322}\en Since
$\{\alpha_{\underline{k}}^{(L)}\}\in l^2({\Bbb{Z}}^2)$ for all
$L>1$, and since
$\frac{1}{(2\pi)^2}\int_0^{2\pi}dP_1\int_0^{2\pi}dP_2\,|G_L(\underline{P})|^2=
\sum_{\underline{k}\,\in{\Bbb{Z}}^2}\,|\alpha_{\underline{k}}^{(L)}|^2$,
we see that $G_L(\underline{P})\in\Lc^2([0,2\pi[\times[0,2\pi[)$.
For this reason there is no problem in recovering the coefficients
$\alpha_{\underline{k}}^{(L)}$ as usual:
$$
\alpha_{\underline{k}}^{(L)}=\frac{1}{(2\pi)^2}\int_0^{2\pi}dP_1\int_0^{2\pi}dP_2
\,G_L(\underline{P})\,e^{-i\underline{P}\cdot\underline{k}}=\frac{1}{(2\pi)^2}\int_0^{2\pi}dP_1\int_0^{2\pi}dP_2
\,\sqrt{F_L(\underline{P})}\,e^{-i\underline{P}\cdot\underline{k}},
$$
with a particular choice of phase. Of course we can repeat our
analysis of the asymptotic behavior of the
$\alpha_{\underline{k}}^{(L)}$'s even now: what we get, using the
same arguments, is that also the sequence
$\{\alpha_{\underline{k}}^{(L)}\}$ decreases to zero for
$\|\underline{k}\|$ diverging faster than any inverse power.

\vspace{1mm}

Moreover we can also check that the following sum rule is satisfied:
\be
\sum_{\underline{k}\,\in{\Bbb{Z}}^2}\,\overline{\alpha_{\underline{k}}^{(L)}}\,c_{\underline{k}}^{(L)}=1,\label{323}\en
for any $L>1$. The proof of this equation makes use twice of the
Poisson summation rule. We have
$$
\sum_{\underline{k}\,\in{\Bbb{Z}}^2}\,\overline{\alpha_{\underline{k}}^{(L)}}\,c_{\underline{k}}^{(L)}=
\frac{1}{(2\pi)^4}\int_0^{2\pi}dP_1\int_0^{2\pi}dP_2\int_0^{2\pi}dQ_1\int_0^{2\pi}dQ_2\,
\sqrt{\frac{F_L(\underline{P})}{F_L(\underline{Q})}}\,\sum_{\underline{k}\,\in{\Bbb{Z}}^2}\,e^{i
(\underline{P}-\underline{Q})\cdot\underline{l}}=
$$
$$
=\frac{1}{(2\pi)^2}\int_0^{2\pi}dP_1\int_0^{2\pi}dP_2\int_0^{2\pi}dQ_1\int_0^{2\pi}dQ_2\,
\sqrt{\frac{F_L(\underline{P})}{F_L(\underline{Q})}}\,\sum_{\underline{k}\,\in{\Bbb{Z}}^2}\,\delta(P_1-Q_1-2\pi
l_1)\,\delta(P_2-Q_2-2\pi l_2).
$$
Now, since $P_j,Q_j\in[0,2\pi[$, the two delta functions reduce to
$\delta(P_j-Q_j-2\pi l_j)=\delta(P_j-Q_j)\,\delta_{l_j,0}$,
$j=1,2$. Therefore we get
$$
\sum_{\underline{k}\,\in{\Bbb{Z}}^2}\,\overline{\alpha_{\underline{k}}^{(L)}}\,c_{\underline{k}}^{(L)}=
\frac{1}{(2\pi)^2}\int_0^{2\pi}dP_1\int_0^{2\pi}dP_2\int_0^{2\pi}dQ_1\int_0^{2\pi}dQ_2\,
\sqrt{\frac{F_L(\underline{P})}{F_L(\underline{Q})}}\delta(P_1-Q_1)\,\delta(P_2-Q_2)=1,$$
as we had to prove.

\vspace{2mm}

Let us now continue the analysis of the consequences of our
orthonormalization procedure considering more in details the
special features of a set of coherent states: which properties of
the set $\C^{(L)}$ can still be proved for $\E^{(L)}$?

The first obvious result is that both these sets produce a
resolution of the identity:
$\sum_{\underline{k}\in{\Bbb{Z}}^2}|\varphi_{\underline{k}}^{(L)}><\varphi_{\underline{k}}^{(L)}|=
\sum_{\underline{k}\in{\Bbb{Z}}^2}|\Psi_{\underline{k}}^{(L)}><\Psi_{\underline{k}}^{(L)}|=\Id_{h_L}$,
where $\Id_{h_L}$ is the identity operator on $h_L$.

Further, let us define the operator $B_L:=X_LbX_L^{-1}$. It is not
hard to check that each $\Psi_{\underline{n}}^{(L)}$ belongs to
the domain of $B_L$. More than this, we can check that
$\Psi_{\underline{n}}^{(L)}$ is an eigenstate of $B_L$ with
eigenvalue $z_{\underline{n}}$. Indeed we have \be
B_L\Psi_{\underline{n}}^{(L)}=\left(X_LbX_L^{-1}\right)\left(X_L\varphi_{\underline{n}}^{(L)}\right)=X_Lb
\varphi_{\underline{n}}^{(L)}=z_{\underline{n}}X_L\varphi_{\underline{n}}^{(L)}=z_{\underline{n}}
\Psi_{\underline{n}}^{(L)}. \label{324}\en

It is easy to compute the commutation rule between $B_L$ and its
adjoint. We get $[B_L,B_L^\dagger]=X_LbX_L^{-2}b^\dagger X_L-
X_L^{-1}b^\dagger X_L^{2}b X_L^{-1}$, which shows that in general
$B_L$ is not an annihilation operator. This is not surprising and,
actually, cannot be avoided since, if $B_L$ were a bosonic
annihilation operator, its eigenstates
$\Psi_{\underline{n}}^{(L)}$ should have surely been not mutually
orthogonal!

\vspace{2mm} It is a well known fact that coherent states saturate
the Heisenberg uncertainty relation $(\Delta \hat q)(\Delta \hat
p)=\frac{1}{2}$. Indeed we easily find $\Delta \hat q=\Delta \hat
p=\frac{1}{\sqrt{2}}$. We ask here if the same is true also for
the vectors $\Psi_{\underline{n}}^{(L)}$. The computation, say, of
$\Delta \hat q$ is not very hard but surely requires some care and
one can check that $(\Delta \hat q)(\Delta \hat p)=\frac{1}{2}$
does not hold. This is not surprising, since the position and
momentum operators do not play such a central role here as for the
{\em canonical} coherent states. For this reason, it is surely
more interesting  to introduce a new operator $Q_L$ which mimics
$\hat q$ in the following sense: since $\hat
q=\frac{b+b^\dagger}{\sqrt{2}}$, and since $b$ has been replaced
by $B_L$ in (\ref{324}), then we put
$Q_L=\frac{B_L+B_L^\dagger}{\sqrt{2}}$. It is now a trivial
computation to check that
$$
(\Delta Q_L)^2=<\Psi_{\underline{n}}^{(L)}, Q_L^2
\Psi_{\underline{n}}^{(L)}>-<\Psi_{\underline{n}}^{(L)}, Q_L
\Psi_{\underline{n}}^{(L)}>^2=\frac{1}{2}\left(\|B_L^\dagger
\Psi_{\underline{n}}^{(L)} \|^2-|z_{\underline{n}}|^2\right),
$$
which would give $1/2$, as in the standard situation, if we had
$X_L=\Id$. In the same way, putting
$P_L=i\frac{B_L^\dagger-B_L}{\sqrt{2}}$ in analogy to $\hat
p=i\frac{b^\dagger-b}{\sqrt{2}}$, we find that
$$
(\Delta P_L)^2=<\Psi_{\underline{n}}^{(L)}, P_L^2
\Psi_{\underline{n}}^{(L)}>-<\Psi_{\underline{n}}^{(L)}, P_L
\Psi_{\underline{n}}^{(L)}>^2=\frac{1}{2}\left(\|B_L^\dagger
\Psi_{\underline{n}}^{(L)} \|^2-|z_{\underline{n}}|^2\right).
$$
Therefore $(\Delta Q_L)(\Delta P_L)=\frac{1}{2}\left(\|B_L^\dagger
\Psi_{\underline{n}}^{(L)} \|^2-|z_{\underline{n}}|^2\right)$,
which is equal to $1/2$ if $X_L=\Id$ but not in general. This is
in agreement with the fact that $[Q_L,P_L]\neq i\Id$. Moreover, it
is not difficult to check that each $\Psi_{\underline{n}}^{(L)}$
saturates again the Heisenberg uncertainty relation in the sense
that, using $[Q_L,P_L]=i[B_L,B_L^\dagger]$, the following equality
holds:  $\Delta Q_L\cdot\Delta
P_L=\frac{1}{2}<\Psi_{\underline{n}}^{(L)},[B_L,B_L^\dagger]
\Psi_{\underline{n}}^{(L)}>$.

\vspace{2mm}

It is now interesting to use our generalized
$(k,q)$-representation to deduce, in analogy with \cite{bgz}, how
should a function  look like in order to produce, together with
its translated, an orthonormal set. In other words, let
$\Psi_{\underline{n}}^{(L)}$ be our o.n. set:
$<\Psi_{\underline{n}}^{(L)},\Psi_{\underline{0}}^{(L)}>=\delta_{\underline{n},\underline{0}}$.
Then we have, inserting the identity operator in (\ref{211}), and
in analogy with what has been done in Proposition 1,
$$
\delta_{\underline{n},\underline{0}}=\int\int_{\Box^{(A)}}
<\Psi_{\underline{n}}^{(L)},\Phi_{k,q}^{(A,a)}><\Phi_{k,q}^{(A,a)},\Psi_{\underline{0}}^{(L)}>\,dk\,dq=$$
$$=\int\int_{\Box^{(A)}}
e^{iqan_1-ikAn_2}\left|<\Phi_{k,q}^{(A,a)},\Psi_{\underline{0}}^{(L)}>\right|^2\,dk\,dq,
$$
which has  $L$ different solutions, i.e. all the functions \be
\left|<\Phi_{k,q}^{(A,a)},\Psi_{\underline{0}}^{(L)}>\right|^2=\left\{
\begin{array}{ll}
\frac{AL}{2\pi aj}, &\mbox{a.e. for} (k,q)\in\left[0,\frac{2\pi}{A}\right[\times\left[0,\frac{aj}{L}\right[,  \\
0, &\mbox{otherwise in } \Box^{(A)},
\end{array}
\right.\label{324bis}\en where $j=1,2,\ldots,L$. In particular, if
$L=1$, then $j=1$ and if $a=A=\sqrt{2\pi}$ we recover the same
result as in \cite{bgz}: in this case
$<\Phi_{k,q}^{(A,a)},\Psi_{\underline{0}}^{(L)}>$ must be a constant
times a phase.

Of course, once we fix the form of
$<\Phi_{k,q}^{(A,a)},\Psi_{\underline{0}}^{(L)}>$, we can recover
the expression of $\Psi_{\underline{0}}^{(L)}$ as a vector in
$h_L$ using the following reconstruction formula
$\Psi_{\underline{0}}^{(L)}=\int\int_{\Box^{(A)}}dk\,dq\,<\Phi_{k,q}^{(A,a)},\Psi_{\underline{0}}^{(L)}>\,
\Phi_{k,q}^{(A,a)}$. A natural question would be to relate the
above solutions of the {\em ortogonality requirement } as obtained
directly using the $(k,q)$-representation with the particular
$\Psi_{\underline{0}}^{(L)}$ we have constructed in (\ref{37}).
This will be done elsewhere.

\vspace{4mm}

We dedicate the last part of this subsection to some perturbative
results concerning our problem starting with an approximated
expression for the coefficients $c_{\underline{n}}^{(L)}$ of the
expansion (\ref{37}). Since
$F_L(\underline{P})=1+F_L^o(\underline{P})$, with
$F_L^o(\underline{P})=\sum_{\underline{m}\in{\Bbb{Z}}^2\setminus{(0,0)}}\,(-1)^{L\,m_1\,m_2}\,e^{-\frac{\pi}{2}\,L(m_1^2
+m_2^2)}\,e^{i\underline{P}\cdot\underline{m}}$, equation
(\ref{314}) can be rewritten as follows: $$
c_{\underline{k}}^{(L)}=\frac{1}{(2\pi)^2}\,\int_0^{2\pi}\int_0^{2\pi}\,\frac{e^{-i\underline{P}
\cdot\underline{k}}}{\sqrt{1+F_L^o(\underline{P})}}\,d\underline{P}=$$
$$=
\frac{1}{(2\pi)^2}\,\int_0^{2\pi}\int_0^{2\pi}\,
e^{-i\underline{P}\cdot\underline{k}}
\left(1-\frac{1}{2}\,F_L^o(\underline{P})+\frac{3}{8}\,F_L^o(\underline{P})^2
+\ldots\right)\,d\underline{P}. $$ Considering only the first two
contributions of this expansion we easily get \be
c_{\underline{k}}^{(L)}\simeq
\delta_{\underline{k},\,\underline{0}}-\frac{1}{2}\left(1-\delta_{\underline{k},\,\underline{0}}\right)\,
(-1)^{Lk_1k_2}\,e^{-\frac{\pi}{2}L(k_1^2+k_2^2)}\label{325}\en Of
course, in order for this approximation to be meaningful, we
further need to restrict ourselves to those $\underline k$ such
that $\underline k=(\pm1,0), (0,\pm 1)$. In fact, a contribution
like $\underline k=(\pm 1,\pm1)$ can only be considered in the
expansion above if we also keep into account those contributions
arising from $\frac{3}{8}\,F_L^o(\underline{P})^2$, which contains
terms of the same order. On the contrary, all these contributions
will be neglected here. Nevertheless we will see that this
apparently rude approximation already produces very good results.
If we introduce the following subset of ${\Bbb{Z}}^2$,
$\Gamma:=\{(1,0),(-1,0),(0,1),(0,-1)\}$, then we get the following
expression for $\Psi_{\underline{n}}^{(L)}$: \be
\Psi_{\underline{n}}^{(L)}\simeq
\varphi_{\underline{n}}^{(L)}-\frac{1}{2}\,e^{-\frac{\pi}{2}L}\sum_{\underline{s}\in\Gamma}
\varphi_{\underline{n}+\underline{s}}^{(L)} \label{326}\en It is
easy to check now that the set of the approximated vectors
$\Psi_{\underline{n}}^{(L)}$ obtained in this way are mutually
orthogonal and normalized with a very good approximation already
for $L=2$. Indeed we find, first of all,
$$\|\Psi_{\underline{0}}^{(L)}\|^2\simeq 1-3e^{-\pi L}=\left\{
\begin{array}{ll}
0.99440, \quad\mbox{if }L=2,  \\
0.99976, \quad\mbox{if }L=3,  \\
0.99999, \quad\mbox{if }L=4,
\end{array}
\right.$$ and so on. Of course, since
$\Psi_{\underline{n}}^{(L)}=T_1^{n_1}T_2^{n_2}\Psi_{\underline{0}}^{(L)}$,
the same norms are obtained for
$\|\Psi_{\underline{n}}^{(L)}\|^2$, $\forall
\,\underline{n}\,\in{\Bbb{Z}}^2$. Moreover, if we compute the
overlap between two neighboring vectors, for instance between
$\Psi_{\underline{0}}^{(L)}$ and $\Psi_{(1,0)}^{(L)}$, we find
that
$$
|<\Psi_{\underline{0}}^{(L)}, \Psi_{(1,0)}^{(L)}>|\simeq\left\{
\begin{array}{ll}
0.00016, \quad\mbox{if }L=2,  \\
0.000001, \quad\mbox{if }L=3,
\end{array}
\right.
$$
and so on. We see that the approximation considered here, which as
we have already remarked looks quite rude, allows to recover
normalization and orthogonalization of the vectors with a
meaningless error already for $L=2$, i.e. for $a^2=4\pi$.
Therefore, we can safely claim that in this way we get a rather
good approximation!

As for the operator $X_L$ and $X_L^{-1}$ we find that $X_L\simeq
\Id-\frac{1}{2}e^{-\frac{\pi}{2}L}\sum_{\underline{s}\in\Gamma}\,T_1^{s_1}T_2^{s_2}$
and $X_L^{-1}\simeq
\Id+\frac{1}{2}e^{-\frac{\pi}{2}L}\sum_{\underline{s}\in\Gamma}\,T_1^{s_1}T_2^{s_2}$
or, more explicitly, \be \left\{
\begin{array}{ll}
X_L\simeq \Id-\frac{1}{2}e^{-\frac{\pi}{2}L}K_L,  \\
X_L^{-1}\simeq \Id+\frac{1}{2}e^{-\frac{\pi}{2}L}K_L,
\quad\mbox{where}\\
K_L=T_1+T_1^{-1}+T_2+T_2^{-1}.
\end{array}
\right. \label{327}\en In order to check that $X_L^{-1}$ above is a
good approximation of the inverse of $X_L$ it is enough to observe
that
$$\|X_LX_L^{-1}-\Id\|=\|X_L^{-1}X_L-\Id\|\leq 4e^{-\pi L}=\left\{
\begin{array}{ll}
0.00747, \quad\mbox{if }L=2,  \\
0.00032, \quad\mbox{if }L=3,
\end{array}
\right.$$ and so on.

\vspace{2mm}

{\bf Remark:} from the above estimates it is clear that the only
{\em dangerous} case is $L=1$, which in fact has not even be
considered. Just as an example, if $L=1$ then we can only prove that
$\|X_1X_1^{-1}-\Id\|\leq 0.17285$, which is surely not enough to
claim that $X_1^{-1}$ as given in (\ref{327}) can be really be
interpreted as the inverse of $X_1$. We will came back on the
situation for $L=1$ shortly.

\vspace{1mm}

Using the expansion (\ref{327}) it is finally possible to derive an
approximated version for $B_L$, which looks now as $B_L\simeq
b-\frac{1}{2}e^{-\frac{\pi}{2}L}[K_L,b]$, so that we get
$$
[B_L,B_L^\dagger]\simeq
\Id-\frac{1}{2}e^{-\frac{\pi}{2}L}\left([[K_L,b],b^\dagger]+[b,[b^\dagger,K_L]]\right),
$$
which {\em converges} toward the identity operator as $L$ diverges,
as expected.

\vspace{2mm}

{\bf Remark:} These o.n. vectors can be used to define to define
certain traces on the von Neumann algebra ${\cal M}_L={\cal
B}(h_L)$. Let ${\cal M}_L^+$ be  the positive part of ${\cal M}_L$.
Then, if we put $\omega_L(X)=\sum_{\underline{n}\in\Bbb{Z}^2} {
<\Psi_{\underline{n}}^{(L)},X\Psi_{\underline{n}}^{(L)}>}$, for
$X\in{\cal M}_L^+$ and  $L>1$,
 this is a faithful  normal trace on ${\cal
 M}_L^+$.

To prove this claim we start noticing that $\omega_L$ is linear.
Moreover, since  $\omega_L(X)$ is a sum of only non negative terms,
the summation and the supremum can be interchanged so that the
normality of $\omega_L$ follows that of each
$<\Psi_{\underline{n}}^{(L)},.\,\Psi_{\underline{n}}^{(L)}>$.

We now prove that $\omega_L(X^*X)=\omega_L(XX^*)$, $X\in{\cal
M}_L^+.$ Indeed we have
$$\omega_L(X^*X)=\sum_{\underline{n}\in\Bbb{Z}^2} {
<\Psi_{\underline{n}}^{(L)},X^*X\Psi_{\underline{n}}^{(L)}>} =
\sum_{\underline{n}\in\Bbb{Z}^2}\|X\Psi_{\underline{n}}^{(L)}\|^2
 = \sum_{\underline{k}\in\Bbb{Z}^2}
\sum_{\underline{n}\in\Bbb{Z}^2}|{<X\Psi_{\underline{n}}^{(L)},\Psi_{\underline{k}}^{(L)}>}|
=$$
$$=\sum_{\underline{k}\in\Bbb{Z}^2}
\sum_{\underline{n}\in\Bbb{Z}^2}|{<X^*\Psi_{\underline{k}}^{(L)},\Psi_{\underline{n}}^{(L)}>}|
=
\sum_{\underline{k}\in\Bbb{Z}^2}\|X^*\Psi_{\underline{k}}^{(L)}\|^2=
\sum_{\underline{k}\in\Bbb{Z}^2} {
<\Psi_{\underline{k}}^{(L)},XX^*\Psi_{\underline{k}}^{(L)}>} =
\omega_L(XX^*)
$$

Moreover, let us suppose that $
0=\omega_L(X)=\sum_{\underline{n}\in\Bbb{Z}^2}\|X^{1/2}\Psi_{\underline{n}}^{(L)}\|^2$,
$X\in{\cal M}_L^+$. Therefore $X=0$, which implies that $\omega_L$
is faithful.



It is finally clear that these considerations can be extended with
no particular difficulty to the general settings introduced in
Section II, but this extension will not be repeated here.

\subsection{The case $L=1$}

We have already noticed that, if $L=1$, the perturbation results
stated above are likely not to work as we would like. This claim
can be actually proven  by the following   {\em reductio ad
absurdum} argument. Suppose that the same procedure discussed
previously also works for $L=1$, so that an o.n. set
$\{\Psi_{\underline{n}}^{(1)}\}$ can be constructed in $h_1=\Hil$.
Let $S$ be the following operator:
$Sf=\sum_{\underline{n}\in{\Bbb{Z}}^2}<\Psi_{\underline{n}}^{(1)},f>\,
\Psi_{\underline{n}}^{(1)}$. It is possible to check that
$S=\sum_{\underline{l},\,\underline{s}\in{\Bbb{Z}}^2}\,
c_{\underline{l}}^{(1)}\overline{c_{\underline{s}}^{(1)}}\,T_1^{l_1-s_1}\,T_2^{l_2-s_2}=X_1^2$,
see (\ref{318}). Indeed using definition (\ref{37}), since $S$ is
bounded and therefore continuous we have, $\forall f,g\in\Hil$,
$$
<f,Sg>=\sum_{\underline{n}\in{\Bbb{Z}}^2}<f,\Psi_{\underline{n}}^{(1)}><\Psi_{\underline{n}}^{(1)},g>=
\sum_{\underline{n},\,\underline{l},\,\underline{s}\in{\Bbb{Z}}^2}\,c_{\underline{l}}^{(1)}
\,\overline{c_{\underline{s}}^{(1)}}<f,\varphi_{\underline{l}+\underline{n}}^{(1)}>
<\varphi_{\underline{s}+\underline{n}}^{(1)},g>=
$$
$$
=\sum_{\underline{n},\,\underline{l},\,\underline{s}\in{\Bbb{Z}}^2}\,c_{\underline{l}}^{(1)}
\,\overline{c_{\underline{s}}^{(1)}}<T_1^{-l_1}T_2^{-l_2}f,\varphi_{\underline{n}}^{(1)}>
<\varphi_{\underline{n}}^{(1)},T_1^{s_1}T_2^{s_2}g>=$$
$$=\sum_{\underline{l},\,\underline{s}\in{\Bbb{Z}}^2}
\,c_{\underline{l}}^{(1)}
\,\overline{c_{\underline{s}}^{(1)}}<T_1^{-l_1}T_2^{-l_2}f,T_1^{s_1}T_2^{s_2}g>,
$$
since $\{\varphi_{\underline{n}}^{(1)}\}$ is complete in $\Hil$.
Therefore $S=\sum_{\underline{l},\,\underline{s}\in{\Bbb{Z}}^2}\,
c_{\underline{l}}^{(1)}\overline{c_{\underline{s}}^{(1)}}\,T_1^{l_1-s_1}\,T_2^{l_2-s_2}$
and, due to (\ref{318}), $S=X_1^2$. Now, since
$\Psi_{\underline{n}}^{(1)}=X_1\varphi_{\underline{n}}^{(1)}$, we
have
$$
\delta_{\underline{n},\,\underline{0}}=<\Psi_{\underline{n}}^{(1)},\Psi_{\underline{0}}^{(1)}>=
<X_1\varphi_{\underline{n}}^{(1)},X_1\varphi_{\underline{0}}^{(1)}>=<S\varphi_{\underline{n}}^{(1)},
\varphi_{\underline{0}}^{(1)}>.
$$
Of course, if the set $\E^{(1)}$ were complete, then we should
have $S=\Id$, which, as the above equality shows, would also imply
that $<\varphi_{\underline{n}}^{(1)},
\varphi_{\underline{0}}^{(1)}>=\delta_{\underline{n},\,\underline{0}}$,
which is false. Therefore the same procedure developed for $L>1$
cannot work for $L=1$!


\section{More difficulties and outcome}

It is very easy to imagine how to extend the procedure described
so far to $\N_N$ for $N>2$, at least if the different unitary
operators commute as for coherent states. More difficult and still
under consideration is the situation when the various $A_j$'s do
not commute. In this case, which is a relevant case, there is
still work to do. We want to close the paper with a couple of such
examples and the difficulties which arise in this case.

The first example we want to mention generalizes that of coherent
states in the following way: the two unitary operators
$T_1=e^{ia\hat q}$ and $T_2=e^{-ia\hat p}$ in (\ref{31}) are now
supposed to satisfy $a^2\neq 2\pi\,L$, for any $L\in\Bbb{Z}$, so
that $[T_1,T_2]\neq 0$. However the two operators can be commuted
paying the price of adding a phase: $T_1\,T_2=T_2\,T_1\,e^{ia^2}$,
and therefore
$T_1^{n_1}\,T_2^{n_2}=T_2^{n_2}\,T_1^{n_1}\,e^{ia^2n_1n_2}$ for all
integers $n_1$ and $n_2$. We can think of repeating the same
procedure, so that we put
$f_{\underline{n}}(x)=T_1^{n_1}\,T_2^{n_2}f_{\underline{0}}(x)$, for
a fixed function $f_{\underline{0}}(x)$ in $\Lc^{2}(\Bbb{R})$, and
then, if $<f_{\underline{n}},f_{\underline{k}}>\neq
\delta_{\underline{n},\underline{k}}$, we define a new function
$\varphi_{\underline{0}}(x)$ as the usual linear combination of the
$f_{\underline{n}}(x)$:
$\varphi_{\underline{0}}(x)=\sum_{k\in{\Bbb{Z}}}\,c_{\underline{k}}\,f_{\underline{k}}(x)$.
We try to fix the expression of the coefficients $c_{\underline{n}}$
by the usual orthonormalization requirement:
$<\varphi_{\underline{n}},\varphi_{\underline{0}}>=\delta_{\underline{n},\underline{0}}$,
where
$\varphi_{\underline{n}}=T_1^{n_1}T_2^{n_2}\varphi_{\underline{0}}$.
The difficulty now arises: equation (\ref{311}) must now be replaced
by the following equation
$$<\varphi_{\underline{n}},\varphi_{\underline{0}}>=\sum_{\underline{k},\,\underline{s}\in{\Bbb{Z}}^2}
\,\overline{c_{\underline{k}}}\,c_{\underline{l}}\,I_{\underline{n}+\underline{k}-\underline{l}}\,
e^{ia^2((n_1-l_1)l_2-(n_2-l_2)k_1)}= \delta_{\underline{n},
\underline{0}}.$$  This is a system of equations, one for each value
of $\underline{n}\in{\Bbb{Z}}^2$, which cannot be solved with the
same strategy adopted to solve  equation (\ref{311}) because of the
presence of the phase $e^{ia^2(\ldots)}$ which makes it impossible
to separate the contributions arising from the $c_{\underline{n}}$
from those arising from $I_{\underline{n}}$.

\vspace{2mm}

The same difficulties also arise in a different context, i.e. when
applying this procedure to a family of non orthogonal wavelets. More
in details, let $T$ and $D$ be the usual translation and dilation
operators acting on a general function $f(x)\in\Lc^2(\Bbb{R})$ as
follows: $(Tf)(x)=f(x-1)$, $(Df)(x)=\sqrt{2}\,f(2x)$. This means,
first of all, that $TD=DT^2$. Let now $f_{\underline{0}}(x)$ be a
fixed function normalized in $\Lc^2(\Bbb{R})$ and suppose that the
various functions
$f_{\underline{l}}(x)=D^{l_1}T^{l_2}f_{\underline{0}}(x)$ are not
mutually orthogonal. We can define a new square integrable function
$\varphi_{\underline{0}}(x)=\sum_{k\in{\Bbb{Z}}}\,c_{\underline{k}}\,f_{\underline{k}}(x)$
and, from this,
$\varphi_{\underline{n}}(x)=D^{n_1}T^{n_2}\varphi_{\underline{0}}(x)$,
$\underline{n}\in{\Bbb{Z}}^2$. The main idea is the usual one: we
try to fix the coefficients of the expansion, $c_{\underline{n}}$,
by requiring that
$<\varphi_{\underline{n}},\varphi_{\underline{0}}>=\delta_{\underline{n},\,\underline{0}}$.
Again: this procedure does not seem to work properly since, even if
we can find an infinite number of equations involving the
$c_{\underline{n}}$'s, again we are not able to solve {\em easily}
this system.

\vspace{2mm}

The conclusion of this short analysis suggests that our procedure,
which works very well when the unitary operators in (\ref{II1})
commute, should be properly generalized when these operators do not
commute! This is exactly our future task and we hope to be able to
solve this problem shortly.

\section*{Acknowledgements}

This work has been financially supported by M.U.R.S.T.


 \appendix

\renewcommand{\theequation}{\Alph{section}.\arabic{equation}}

 \section{\hspace{-.7cm}ppendix:  Generalized kq-representation}

The relevance of the $kq$-representation in many-body physics has
been extabilished since its first appearances, \cite{zak}. What was
originally a physical tool has became, during the years, also a
mathematical interesting object, widely analyzed in the literature,
see \cite{dau2,jan} for instance. We give here only few definitions
and refer to \cite{zak,jan,zak2} and \cite{bgz} for further reading
and for applications.

The origin of the $kq$-representation consists in the well known
possibility of a simultaneous diagonalization of  two commuting
operators. In \cite{zak2} it is shown that the following
distributions \be \psi_{kq}(x)=\sqrt{\frac{2\pi}{a}}
\sum_{n\in\Bbb{Z}}e^{ikna}\delta(x-q-na), \quad\quad k\in[0,a[,
\quad q\in\left[0,\frac{2\pi}{a}\right[ \label{26} \en are
(generalized) eigenstates of both $T(a)=e^{ipa}$ and
$\tau(\frac{2\pi}{a})=e^{ix2\pi/a}$. Here $a$ is a positive real
number which plays the role of a lattice spacing.

As discussed in \cite{zak2}, these $\psi_{kq}(x)$ are Bloch-like
functions corresponding to infinitely localized Wannier functions.
They also satisfy orthogonality and closure properties. This implies
that, roughly speaking, they can be used to define a new
representation of the wave functions by means of the integral
transform $Z: \Lc^2(\Bbb{R})\rightarrow \Lc^2(\Box)$, where
$\Box=[0,a[\times[0,\frac{2\pi}{a}[$, defined as follows: \be
h(k,q):=(ZH)(k,q):=\int_{\Bbb{R}}d\omega
\overline{\psi_{kq}(\omega)}H(\omega), \label{27} \en for all
functions $H(\omega)\in \Lc^2(\Bbb{R})$. The result is a function
$h(k,q)\in \Lc^2(\Box)$.

To be more rigorous, $Z$ should be defined first on the functions of
${\cal C}_o^\infty(\Bbb{R})$ and then extended to $\Lc^2(\Bbb{R})$
using its continuity, \cite{jan}. In this way it is possible to give
a rigorous meaning to formula (\ref{27}) above. In most
applications, \cite{antbag2}, the {\em lattice spacing} $a$ is
chosen as $a^2=2\pi$. Here we are  interested in a more general
situation: we need to consider a different lattice with rectangular
lattice cells with surface $2\pi L$, $L=1,2,3,\ldots$.

\vspace{2mm}

Let therefore $T(a)=e^{i\hat p a}$ and $\tau(b)=e^{i\hat q b}$,
with $ab=2\pi L$, for some natural $L$. It is clear that for all
possible $L\in\Bbb{N}$ the two operators still commute:
$[T(a),\tau(b)]=0$. For each given $A>0$ let us define the set of
(generalized) functions \be\Phi_{k,q}^{(A,a)}(x)=
\sqrt{\frac{A}{2\pi}}\sum_{l\in\Bbb{Z}}\,e^{iklA}\,\delta(x-q-la),\label{28}\en
where
$(k,q)\in\Box^{(A)}:=\left[0,\frac{2\pi}{A}\right[\times[0,a[$. If
$\xi_x$ is the generalized eigenvector of the position operator
$\hat q$, $\hat q\xi_x=x\xi_x$, we write $\Phi_{k,q}^{(A,a)}(x)$
as $\Phi_{k,q}^{(A,a)}(x)=<\xi_x,\Phi_{k,q}^{(A,a)}>$.

Is not hard to prove the following statements:
\begin{enumerate}
\item \be T(a)\Phi_{k,q}^{(A,a)}(x)=e^{ikA}\Phi_{k,q}^{(A,a)}(x), \quad \tau(b)
\Phi_{k,q}^{(A,a)}(x)=e^{iqb}\Phi_{k,q}^{(A,a)}(x),\label{29}\en
\item
\be\int\int_{\Box^{(A)}}\overline{\Phi_{k,q}^{(A,a)}(x)}\,\Phi_{k,q}^{(A,a)}(x')\,dk\,dq=\delta(x-x'),\label{210}\en
\item
\be\int\int_{\Box^{(A)}}|\Phi_{k,q}^{(A,a)}><\Phi_{k,q}^{(A,a)}|=\Id,\label{211}\en
where the usual Dirac bra-ket notation has been adopted;
\item \be\int_{\Bbb{R}}\overline{\Phi_{k,q}^{(A,a)}(x)}\,\Phi_{k',q'}^{(A,a)}(x)
\,dx=\delta(k-k')\,\delta(q-q').\label{212}\en
\end{enumerate}

The proof of these statements does not differ significantly from
the standard one, and will be omitted here. We just want to remark
that, for general $a$ and $a'$, we find that
$T(a)\Phi_{k,q}^{(A,a')}\neq e^{ikA}\Phi_{k,q}^{(A,a')}(x)$. In
other words, in general $\Phi_{k,q}^{(A,a')}(x)$ is not an
eigenstate of $T(a)$ if $a\neq a'$.

Also, it should be noticed that the value of the parameter $b$
entering in the definition of $\tau(b)$, is fixed by requiring
that $T$ and $\tau$ commute but play no role in the definition of
the lattice cell $\Box^{(A)}$, which on the other way is defined
by an extra positive parameter, $A$, which needs not to be related
to $b$ itself. However, quite often in applications $A$ coincides
with $a$ and with $b$.

\end{document}